\documentstyle[aps,pre,preprint]{revtex}
\tightenlines
\begin{document}
\draft 
\title{Spectroscopic Constraints on the Surface Magnetic Field
of the Accreting Neutron Star EXO 0748-676} 
\author{Abraham Loeb}
\address{Institute for Advanced Study, Einstein Drive, Princeton, NJ
08540\footnote{On Sabbatical leave from the Astronomy Department,
Harvard University, 60 Garden Street, Cambridge, MA 02138;
aloeb@cfa.harvard.edu}} 
\date{\today} 
\maketitle

\begin{abstract} 

Gravitationally redshifted absorption lines of Fe XXVI, Fe XXV, and O
VIII were inferred recently in the X-ray spectrum of the bursting
neutron star EXO 0748-676.  We place an upper limit on the stellar
magnetic field based on the iron lines.  The oxygen absorption feature
shows a multiple component profile that is consistent with Zeeman
splitting in a magnetic field of $\sim (1$--$2)\times 10^9$G, and for
which the corresponding Zeeman components of the iron lines are
expected to be blended together. In other systems, a field strength
$\gtrsim 5\times 10^{10}$G could induce a blueshift of the line
centroids that would counteract gravitational redshift and complicate
the derivation of constraints on the equation of state of the neutron
star.

\end{abstract}
\input epsf.sty

An unusually long exposure by the XMM-{\it Newton} observatory of
X-ray bursts from the accreting neutron star EXO 0748-676 revealed the
existence of redshifted absorption lines such as the $n=2\rightarrow
3$ transitions of hydrogen-like and helium-like iron and the
$n=1\rightarrow 2$ (Ly$\alpha$) transition of hydrogen-like oxygen
\cite{Cottam}.  The common redshift of the lines, $z_{\rm obs}=0.35$,
is consistent with absorption features from the surface of a neutron
star with a composition of normal nuclear matter\cite{Lattimer}.  The
observed equivalent width of the Fe lines, $\sim 10$eV, is an order of
magnitude larger than expected from thermal Doppler broadening at the
inferred surface temperature of 1--2 keV, and so it may be dominated
by the Stark effect\cite{Bildsten}. Rapid rotation of the neutron star
at a spin frequency of $\sim 10^2$ Hz may contribute substantially to
the observed width of the redshifted lines \cite{Ozel}. Other
identified lines do not show any significant redshift and are likely
to be associated with gas far away from the neutron star surface.

Here we consider the potential contribution of the Zeeman effect to
both the shift and the width of the observed lines. The detectability
of this contribution was first predicted in
Ref. \cite{Bahcall}. Indirect theoretical arguments suggest that the
magnetic field of accreting neutron stars in low-mass X-ray binaries
may have an amplitude of up to $\sim 10^{10}$G
\cite{Cumming,Miller}. Such a field could have important effects on
the accretion flow near the star and on the existence of pulsations in
its emission\cite{Psaltis}. However, no direct measurement of the
field strength has been reported so far.

A relatively weak magnetic field induces a Zeeman (or
Paschen-Back\cite{Paschen}) splitting of an atomic level by an energy
separation which is linear in the magnetic field strength, $\sim (e h/
4\pi m_e c) (M_L +2M_S) B= 5.8~{\rm eV} (M_L+2M_S) \left({B/10^{9}{\rm
G}}\right)$, where $M_L$ and $M_S$ are the quantum numbers of the
orbital angular momentum and the spin of the electron.  The split
transition lines associated with changes of $\Delta M_L =\pm 1$ and
$\Delta M_S=0$ would therefore be maximally separated by an energy,
\begin{equation}
(\Delta E)_{\rm split} \sim 11.6~{\rm eV} 
\left({B\over 10^{9}{\rm G}}\right).
\label{eq:splitting}
\end{equation}
For a strong field, there is a net blueshift of the centroid of the
transition line components which is quadratic in $B$, and for
hydrogen-like ions is given by\cite{Jenkins,Schiff,Preston},
\begin{equation}
\left(\Delta E\right)_{\rm shift} \sim {e^2a_0^2\over 8 Z^2 m_e c^2}
n^4(1+M_L^2)B^2 =9.2\times 10^{-4}~{\rm eV}~\left({Z \over
26}\right)^{-2} n^4(1+M_L^2) \left({B\over 10^{9}{\rm G}}\right)^2,
\label{eq:shift}
\end{equation}
where $n$ and $M_L$ are the principal and orbital quantum numbers of
the upper state, $a_0$ is the Bohr radius, and $Z$ is the nuclear
charge ($=26$ for Fe and 8 for O).  The factor of $Z^{-2}$ was added
to account for the reduction in the square of the orbital radius of
the last bound electron in heavy elements as compared to hydrogen.

Equation (\ref{eq:splitting}) implies that the intrinsic (redshift
corrected) full width at half minimum of the Fe absorption lines in
the neutron star EXO 0748-676 spectrum, $\delta E\sim 20~{\rm eV}$,
limits the strength of its surface magnetic field to a value
$B\lesssim 1.7 \times 10^9$G, assuming that a weak splitting with
$(\Delta E)_{\rm split}<\delta E$ cannot be resolved but is rather
blended by Stark broadening \cite{Bildsten}, rotational
broadening\cite{Ozel}, or the current instrumental sensitivity.  This
is the first direct limit on the magnetic field of a neutron star in a
low mass X-ray binary.  An independent constraint on the field
strength can be derived from the line shift.  Because both the iron
and oxygen features showed the same redshift, $z_{\rm obs}$, for
transitions with different $Z$ and $n$, the Zeeman blueshift, $z_{\rm
B}\equiv -(\Delta E/E)_{\rm shift}$, must be much smaller than the
observed redshift, $\vert z_{\rm B}\vert\ll z_{\rm obs}$.  Equation
(\ref{eq:shift}) then implies $B\ll 5\times 10^{10}$G for the iron and
oxygen lines, a much weaker upper limit than derived from the line
width.  More generally, $(1+z_{\rm obs})=(1+z_{\rm grav})(1+z_{\rm
B})$, or equivalently $z_{\rm grav} =(z_{\rm obs}-z_{\rm B})/(1+z_{\rm
B})$, where $z_{\rm grav}$ is the gravitational redshift. An upper
limit on the magnetic field strength from the line width translates to
a lower limit on the gravitational redshift,
\begin{equation}
z_{\rm grav}\geq {z_{\rm obs} + \vert z_{\rm B,max}\vert \over
1 - \vert z_{\rm B,max}\vert} ,
\end{equation}
where a line with energy $E=hc/\lambda$ and an intrinsic (redshift
corrected) wavelength width $\delta \lambda$, has an intrinsic energy
width of $\delta E=h c\delta\lambda/\lambda^2$ and
\begin{equation}
\vert z_{\rm B,max}\vert \equiv \left( {E\over
920~{\rm eV}}
\right)^{-1} \left({Z\over 26}\right)^{-2}\left({n\over 10}\right)^4
\left(1+M_L^2\right)
\left({{\rm \delta E}\over 116~{\rm eV}}\right)^2 .
\end{equation}
For other neutron stars, constraints on the equation of state of the
neutron star or the theory of gravity\cite{Lattimer,Dedeo} could be
complicated by the possible existence of a magnetic field with a
strength $\gtrsim 5\times 10^{10}(n/3)^{-2}$G.  The magnetic effect
can be isolated based on the particular dependence of the quadratic
Zeeman shift on $n$ and $Z$ in Eq. (\ref{eq:shift}).  While the above
analytical estimates are restricted to moderate fields [$\ll 10^{11}
(Z/26)^{2}$G], a precise evaluation of the same limits for arbitrary
field strength in other systems can be done based on the Tables in
Ref. \cite{Ruder}.

Although no multiple components are clearly seen for the iron lines,
the Ly$\alpha$ absorption feature of hydrogen-like oxygen is split
into multiple (possibly three) components in the observed spectrum
\cite{Cottam}.  If this splitting is caused by the Zeeman effect, it
implies a magnetic field strength of $\sim 1.4\times 10^9$G (including
the redshift correction)\cite{comment}.  If the intrinsic width of the
O VIII line is caused by rotational broadening (requiring a rotation
period of $\sim 10$ millisecond), then for other lines $\delta \lambda
\propto \lambda$ while Zeeman splitting results in a separation
between the outer line components of $(\Delta \lambda)_{\rm
split}\propto \lambda^2$. Scaling in the late burst spectrum from the
outer two O VIII absorption features centered at observed wavelengths
of 25.2\AA ~and 26.4\AA, each with an observed width $\sim 0.53$\AA,
down in wavelength to the Fe XXV absorption feature at 13.75\AA ,
implies a redshifted width $(1+z_{\rm obs})\delta \lambda \sim
0.29$\AA ~that is comparable to the extrapolated Zeeman split
$(1+z_{\rm obs})(\Delta \lambda)_{\rm split}\sim 0.34$\AA ~for the
iron line. Moreover, the Ly$\alpha$ line of oxygen should be split
into a triplet while the Fe$^{24+}$ $2\rightarrow 3$ line should have
many more components (see Table A4.1 in \cite{Ruder}). Hence, we
conclude that {\it the possible Zeeman components of the iron line may
be blended together due to their finite width even though the
splitting of the oxygen Ly$\alpha$ line is resolved}. Zeeman splitting
provides a viable alternative to the other interpretation of the
multi-component, broad profile of oxygen given by Cottam et
al. \cite{Cottam}, involving an extended outer atmosphere with a
possible slow outflow and a self-reversed (W-shaped) absorption
profile. The latter interpretation does not immediately explain why
the W-shaped broad profile does not appear in the iron lines. The
Zeeman interpretation predicts that all redshifted lines of observed
wavelength $\lambda \gtrsim 20$\AA ~would show a multi-component
profile with a wavelength split among its outer components that scales
as $\lambda^2$.

While the above evidence for a magnetic field is only tentative,
future X-ray observations with longer exposure times (or further
analysis and detailed modeling of existing data) can easily test the
generic signatures of a Zeeman effect.  The contributions of the Stark
effect and the rotation of the neutron star to the width of the
lines\cite{Bildsten,Ozel} can be separated by measuring the line width
for transitions with different quantum numbers and line energies. A
magnetic field strength of $\sim (1$--$2)\times 10^9$G is rather
plausible for accreting neutron stars in low mass X-ray binaries,
since these systems are believed to be the progenitors of millisecond
radio pulsars which possess fields of $10^8$--$10^9$G
\cite{Batta}. Future X-ray observations with improved flux sensitivity
and spectral resolution may provide conclusive evidence for the
surface field strength in EXO 0748-676 and other accreting neutron
stars.

\bigskip
\bigskip
\paragraph*{Acknowledgments.}

The author thanks Jean Cottam, Jeremy Heyl, Feryal \"Ozel, Dimitrios
Psaltis, and George Rybicki for useful discussions.  Sabbatical 
support from the Institute for Advanced Study at Princeton and the
John Simon Guggenheim Memorial Fellowship is gratefully acknowledged.
This work was also supported in part by NSF grants AST-0071019,
AST-0204514 and NASA grant NAG 5-13292.


\begin{references}
\bibitem{comment} The inferred field strength refers to an average
over the emission region on the stellar disk. For an analogous
situation in white dwarfs, see chapter 7 in Ref. \cite{Ruder}.  X-ray
polarization measurements within the absorption lines may ultimately
be used to constrain the magnetic field geometry.
\end{references}
\end{document}